# Time-Energy Uncertainty in Neutrino Resonance: Quest for the Limit of Validity of Quantum Mechanics

R. S. Raghavan
Instiiute of Particle, Nuclear and Astronomical Sciences and Department of Physics, Virginia Polytechnic Institute and State University, Blacksburg VA 24061

The role of quantum mechanical time-energy uncertainty (TEU) is central in resonant neutrino ($\tilde{\nu}_e$) reactions $^3H \leftrightarrow ^3He$ because of the unusual 18 y lifetime of $^3H$. The TEU explicitly manifests itself by a non-intuitive but quantitatively predictable spontaneous temporal growth of the $\tilde{\nu}_e$ resonance signal. A *slower* growth rate signifies violation of TEU via a larger than natural width of $^3H$, possibly imposed by a fundamental length in nature. Strong limits on TEU violation can be set in the unprobed virgin energy regime of $\sim 10^{-24}$ eV.

The possibility of resonant nuclear reactions $^3H \leftrightarrow ^3He$ induced by hypersharp (anti)neutrino ($\tilde{\nu}_e$) lines from the 2-body decay of $^3H$ (tritium) has recently been discussed in detail [1]. The long lifetime of the decaying state (~18 years) facilitates critical examination of the meaning, implications and external control of the time-energy uncertainty (TEU) principle, a bed-rock of quantum mechanics. The unusual lifetime has also generated incorrect predictions for the resonance that betray basic misunderstanding of the TEU itself. They need clarification and correction. The TEU plays a central positive role in the tritium $\tilde{\nu}_e$ resonance. Indeed, we predict a novel, non-intuitive effect that explicitly demonstrates the TEU with a spontaneous *temporal growth* of the resonance absorption at zero detuning. The growth rate is quantitatively predictable by the measured $^3H$ *decay* rate if the TEU is valid.

The tritium $\tilde{\nu}_e$ resonance opens further, the opportunity to explore the limits of validity of the TEU, thus, quantum mechanics itself. The $\tilde{\nu}_e$ resonance is reminiscent of γ-ray resonances (Mőssbauer Effect ME). But the long lifetime of tritium and the atomic dynamics of $^3H$ and $^3He$ specific to the experimental framework[1] present entirely new facets well beyond those of classical ME. The significant discovery in (1) is that a *useful fraction* of the tritium $\tilde{\nu}_e$ will always be emitted with the *natural width*, $\hbar/\tau \sim 10^{-24}$ eV in spite of myriad extra-nuclear sources of broadening and energy shifts well known in ME work[2] (see below for brief details). They affect only the intensity, not the *width* as they do in short lived ME levels.

The extreme sharpness, $\Delta E/E \sim 5\times10^{-29}$, opens up new probes of fundamental physics. An example, the focus in this Letter, is the quest for the limits of validity of the TEU at extraordinarily small energies $\sim 10^{-24}$ eV. We show that violation of TEU appears as an anomalously *slower* growth of the resonance signal than that expected from the decay rate of tritium. The effect can arise due to line width broader than the natural width. The signal growth anomaly can set limits on such broadening to a fraction of the natural width. The critical role of the hypersharp standard of energy of the $\tilde{\nu}_e$ resonance is to ensure the fundamental origin of such broadening.

The question is more than academic because it has been long suggested [3] that a Fundamental Length $L = [G\hbar/c^3]^{1/2}$ in nature may impose a limiting width broader than the natural width, thus violating TEU. Involving as L does, the constants of Planck and Newton, an indication or otherwise of TEU violation may provide the first empirical milestone or guide stone on the road to quantum gravity.

The quantum mechanical uncertainty relation $\Delta E\ \Delta t \sim \hbar$ relates the energy width ΔE of a decaying state with the time of observation Δt. If in a measurement, no restriction is placed on Δt, the "time of observation" is the mean lifetime τ of the state. In this case a measurement of the energy of the state shows a spread $\Delta E_{nat}$, the natural width of the state. State widths can be measured via resonance reactions. The maximum reaction cross section $\sigma_o$ (height of the resonance line) is governed by the flux density of γ or $\tilde{\nu}_e$ emitted within the resonance energy width ΔE. In the case of the natural width $\Delta E_{nat}$, $\sigma_o$ takes the maximum value of $\sigma_o = 2\pi\, ƛ^2$ where $ƛ$ is the $\tilde{\nu}_e$ wavelength. In $^3H \leftrightarrow {}^3He$ $\sigma_o$ is very large, $\sim 10^{-17}$ $cm^2$. For line widths broadened beyond $\Delta E_{nat}$, $\sigma_o$ drops inversely as the line width (the total area integrated over the resonance curve is independent of the line width). The value of $\sigma_o$ measured as the percent absorption at zero detuning of the resonance is the experimental handle on the line width in the following.

The basic questions here are: the definition of the "time" $\Delta t$ in the TEU, the resulting energy width $\Delta E$ and the effect on $\sigma_o$. Surprisingly confusing opinions with major experimental consequences have recently been expressed in the literature. Lipkin[4] asserts that the TEU with *his* definition of the "time" would make it impossible to observe the $\tilde{\nu}_e$ resonance. Potzel [2] asserts that the TEU does not play any role in the $\tilde{\nu}_e$ resonance line width while Akhmedov et al[5] conclude that an imprecisely defined "time of the experiment" does. In this Letter we affirm the key role of TEU in the tritium $\tilde{\nu}_e$ resonance but refute Lipkin's assertion on the consequences. Indeed, the effect offers a sensitive tool not only to demonstrate the action of TEU but experimentally test its validity in unprobed energy regimes.

The confusion arises from incorrect definition of the "time of observation". Lipkin *explicitly* defines this as the time interval between $\nu$ (or $\gamma$) emission in the source and its detection. This definition can only be interpreted as the $\tilde{\nu}_e$ time of flight (TOF) which can be deduced from the time point of detection. Then Lipkin asserts that since this TOF is necessarily much smaller than the tritium lifetime (or for that matter, isomeric $\gamma$-ray nuclear lifetimes), the energy spread should be very large in laboratory geometries. As a result the flux density at the resonance maximum is diluted [ $(\delta t_{TOF}/\tau) <<<1$] to the extent that it would be "impossible to observe the resonance" particularly for the tritium $\tilde{\nu}_e$. The argument leads to obvious contradictions since it implies that the measured width depends on the *baseline* of the experiment. If that were the case, the line width of the $^{57}Fe$ ME resonance ($\tau \sim 140$ ns) in a bench top baseline of ~1 cm ($\delta t_{TOF}/\tau \sim 2\times10^{-4}$) would be far broader than that in the Pound-Rebka gravitational red shift experiments[6] with $^{57}Fe$ with a baseline of 2200 cm (($\delta t_{TOF}/\tau \sim 0.5$). *This was not the case*.

The only meaningful definition of "the time" in TEU is the time spent by the level as an unstable object, i.e. the duration T after the *creation of the level* before it decays. If no external attempt is made to control T, the duration is simply the mean lifetime $\tau$ of the level. If on the other hand an explicit determination of T is made, i.e. if the resonance is observed after a *measured* duration $T << \tau$, it interferes in the natural process. Then the TEU predicts a spread in the measured energy $\Delta E > \Delta E_{nat}$.

The TEU effect can be controlled in a fundamental way if the time of *creation* is known, thus the "age" of the decaying level is determined. Just such an experiment was performed using the ME. The 14.4 keV state that emits the $^{57}Fe$ ME line is preceded by a 122 keV $\gamma$-ray that populates the ME level, thus the birth of the state can be signaled by the 122 keV $\gamma$. The ME resonance signal vs. the delay time T (= age of the level) between the 122 and 14.4 keV ME $\gamma$ signals in this "time-filtered resonance" was observe [7] [8]. The results show line broadening that varies inversely as the delay time T and that the consequent decrease of the maximum resonant cross section A at zero detuning, agrees with theoy [9] [10]. Thus:

$$A \propto [1 - J_o^2(\beta\, T/\tau)^{1/2}] \qquad (1).$$

$J_o$ is a Bessel coefficient of zero order. $\beta$ = $Nf\sigma$ the resonance thickness of the absorber in mean free path units. $\sigma = \sigma_o$ if the level width in the absorber is the natural width. T is the the delay time between the 122 and 14.4 keV signals, the "radio-age" of the level. Eq. (1) shows that the absorption A increases with the delay time T as $\Delta E \propto 1/T$ narrows according to the TEU and $\sigma \propto 1/\Delta E$ increases. $\sigma \rightarrow \sigma_o$ as $T/\tau \rightarrow 1$ and $\Delta E \rightarrow \Delta E_{nat}$. The result is a controlled test of the TEU which is thus well established even in short lived ME resonance, refuting Lipkin's assertion.

If we apply these results to the case of the tritium $\tilde{\nu}_e$ resonance, the long lifetime of several years instead of 140 ns makes it straightforward to observe the time filtering effect. The radio-age T of the tritium activity is known since the date of its original activation (e.g. in reactor etc., with activation time short compared to the 18 year lifetime) is known. The tritium radio-age T is thus the controlling "time" in the TEU. Since nuclear decay is normally immune to external influences, it is immaterial whether after creation, the tritium activity (usually a gas) is stored elsewhere in the period before being installed in a solid matrix and counted in the resonance measurement. The radio-age-dependent TEU broadening affects the resonance absorption as in (1) and grows with the radio age of T as shown in Fig. 1. Using the same absorber, the growth curve for older ages (~10s of years) can be obtained with different $^3H$ samples stored for long times after activation.

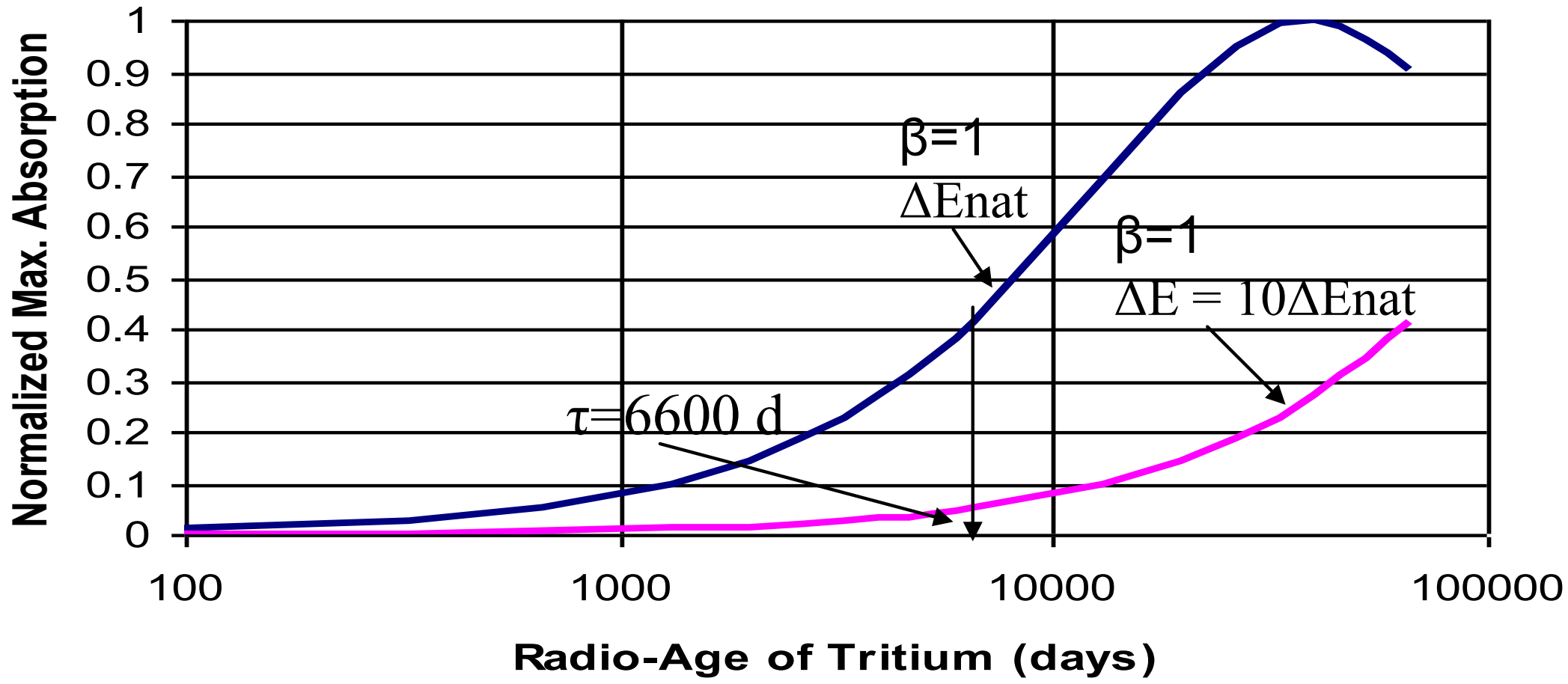

Fig. 1 Growth rates of normalized resonance absorption A at zero detuning (eq. 1) with radio-age of $^3$H for QM-TEU and violation of QM-TEU. Left curve for $\Delta E(QM) = \Delta E_{nat}$, as required by TEU corresponding to the measured $\tau$ of $^3$H for known absorber thickness $\beta$=1. Right curve for hypothetical $\Delta E(nonQM) = 10\Delta E_{nat}$ for $\beta$=1.

Indeed, the source *continues to age every day* during the experiment thus the resonance signal rate in every source *spontaneously increases with time.* The TEU effect is thus directly manifested by signal *growth* rates in a single source (independent of knowledge of the activation date). The novel TEU effect predicted in Fig. 1 is an explicit demonstration of the TEU as overtly compelling as nuclear decay itself.

In the experimental method of resonance $\tilde{\nu}_e$ activation[1] of the $^3$He absorber ($^3$He→ $^3$H) and its detection by the β-emission of the newly produced $^3$H, the actual growth rate is faster. The signal rate grows additionally with the *activation time*, i.e. duration of $\nu ?_e$ exposure. Fig 1 does not include this growth.

The value of β=1 in Fig. 1 indicates a thin resonance absorber of just 1 resonance mean free path. Thus the observed width is determined essentially by the maximum cross section σ included in β which, in turn, depends on the prevailing state width. A basic question is how to ensure that the observed broadening is "fundamental". The width could in principle, be broader due to environmental "solid state" effects.

It is here that the discovery (in ref. 1) of the special nature of the $^3$H↔$^3$He resonance is of key relevance. Extranuclear interactions that create broadening in classical ME in short lived states, if applied blindly, would imply a broadening $\sim 10^{12}$ times the natural width of tritium[2]. Nevertheless the surprising finding in ref. 1 is that a useful fraction of the tritium line is emitted with the natural width.

The reasons for this remarkable new effect are:
1) Considering the ambient fluctuations as a modulation of the central energy (carrier) in the long lifetime of the state (relative to all other relevant times) the perturbations (periodic or stochastic) are averaged out[11]. The result is no broadening but only some loss of carrier *intensity;* the lost intensity is transferred to *side bands* that are pushed out *far away* as a direct result of the long lifetime. A *hypersharp fraction* always remains. For the same perturbations in *short lived* ME states (wide lines) the side bands are close to the carrier so that classical ME lines are routinely *broadened.* The physics of line broadening here is thus crucially different in short (μs) and long lifetimes (yrs).

(2) In every $^3$H decay, both $^3$H *and* the daughter $^3$He participate. As a rule the net final state energy differences of the two atoms cancel exactly with opposite signs in the emission+ absorption process. The resonance condition is thus always strictly maintained.

3) The quadrupole moments of both nuclei are zero thus, distributed electric interactions due to lattice defects are absent.

4) The imbedding of the gaeous $^3$H and $^3$He in metals (as described in ref. 1) results in these atoms being trapped in *local* potential wells. The atomic dynamics of $^3$H and $^3$He, are controlled by *discrete excitations within* the well[12] (observed in neutron scattering [13] in the material proposed for the tritium resonance). The bulk lattice Debye excitation spectrum plays no role as in the ME.

5) Because of (4) the interactions of $^3$H and $^3$He In the potential wells are shielded from inhomogeneties in the bulk lattice.

6) The lowest vibational level in the well corresponds to ~800K which forbids low frequency excitations. The tritium resonance signal is thus *temperature independent* unlike the classical ME.

7) For the same reason temperature dependent energy shifts[14] are also absent.

8) The zero point and excitation energies in the potential well depend on the spatial coordinates of atoms of the trap. These are also motionally averaged to unique central values by vibrations, maintaining the hyper-precise energy balances of (2) above.

In summary, to the extent that the $^3$H↔$^3$He resonance occurs at all, it does with *no line broadening* i.e. with the natural width of the tritium state. This width of ~$10^{-24}$ eV is the new energy standard observable in nature. It provides the basis for a critical test the TEU. Against this standard, a *fundamental* state width ΔE broader than the natural width can arise only by a violation of the TEU. Experimentally, this will be manifested by a reduced cross section σ as $\Delta E_{nat}/\Delta E < 1$. Ther result is an anomalously *slower growth rate* of the signal as shown in the right curve in Fig. 1. A TEU violation due to a state width ΔE(nonTEU)/E ~$10^{-29}$ (a fraction of the basic ΔE/E~ 5x$10^{-29}$ of the $^3$H↔$^3$He resonance), appears easily observable. Stringent limits can thus be set on the validity of the TEU.

The combination $L(cm)=[G\hbar/c^3]^{1/2}$ consisting entirely of universal constants is traceable to Planck (hence Planck Length). With a value of ~$10^{-33}$ cm, it has been interpreted as the "quantum" of length or Fundamental Length. Mead[2] has suggested further that L as a limit on *measurable length* also sets a fundamental *"uncertainty"* in a measured *length* Δ*l*/*l* akin to that in canonical measurables in quantum mechanics. Thus he has argued that L affects energies of nuclei via the imprecision it imposes on r in the nuclear potential V(r). .Mead then predicts a fundamental nuclear *width* that depends on L (not on the TEU) as ΔE/E(L) = (L/R) λ. R is the nuclear radius. λ depends presumably on the quantum gravity model. ΔE/E(L).~ $10^{-20}$ for λ = 1 and ~$10^{-40}$ for λ = (L/R) . A strong limit down to ΔE/E~$10^{-29}$ from the TEU effect in the $^3$H↔$^3$H $\tilde{\nu}_e$ resonance would thus be a significant empirical milestone in the conceptual and quantitative development in theories of quantum gravity.

I thank Patrick Huber for interesting and useful discussions